\documentclass[tradiabstract]{aa}

\usepackage{graphicx}
\usepackage{natbib}
\bibpunct{(}{)}{;}{a}{}{,}
\usepackage{txfonts}

\newcommand{\micron}{\rm \mu m}
\newcommand{\EW}{W_\lambda}
\newcommand{\MBH}{M_{\rm BH}}
\newcommand{\Edd}{\mathrm{Edd}}
\newcommand{\ledd}{l_{\rm Edd}}
\newcommand{\RNLR}{R_{\rm NLR}}
\newcommand{\LX}{L_{\rm X}}
\newcommand{\LMIR}{L_{\rm MIR}}
\newcommand{\Msun}{M_{\sun}}

\begin{document}

\title{Discovery of a strong Baldwin effect in mid-infrared AGN lines\thanks{Based on ESO observing programmes 078.B-0303, 080.B-0240, and the DDT program 280.B-5068.}}

\author{S.~F.~H\"onig\inst{1} \and
A.~Smette\inst{2} \and
T.~Beckert\inst{1} \and
H.~Horst\inst{3,4} \and
W.~Duschl\inst{3} \and
P.~Gandhi\inst{5} \and
M.~Kishimoto\inst{1} \and
G.~Weigelt\inst{1}}

\offprints{S.~F. H\"onig \\ \email{shoenig@mpifr-bonn.mpg.de}}

\institute{
Max-Planck-Institut f\"ur Radioastronomie, Auf dem H\"ugel 69, 53121 Bonn, Germany \and
European Southern Observatory, Casilla 19001, Santiago 19, Chile \and
Institut f\"ur Theoretische Physik und Astrophysik, Christian-Albrechts-Universit\"at zu Kiel, Leibnizstr. 15, 24098 Kiel, Germany \and
Zentrum f\"ur Astronomie, ITA, Universit\"at Heidelberg, Albert-Ueberle-Str. 2, 69120 Heidelberg, Germany \and
RIKEN Cosmic Radiation Lab, 2-1 Hirosawa, Wakoshi Saitama 351-0198, Japan
}

\date{Received 5 May 2008 / Accepted 14 May 2008}

\abstract{We present the discovery of a Baldwin effect in 8 nearby Seyfert galaxies for the three most prominent mid-infrared forbidden emission lines observable from the ground that are commonly found in AGN, [\ion{Ar}{iii}]($\lambda8.99\,\micron$), [\ion{S}{iv}]($\lambda10.51\,\micron$), and [\ion{Ne}{ii}]($\lambda12.81\,\micron$). The observations were carried out using the VLT/VISIR imager and spectroraph at the ESO/Paranal observatory. The bulk of the observed line emission comes from the inner $<$0\farcs4 which corresponds to spatial scales $<$100\,pc in our object sample. The correlation index is approximately $-0.6$ without significant difference among the lines. This is the strongest anti-correlation between line equivalent width and continuum luminosity found so far. In the case of Circinus, we show that despite the use of mid-infrared lines, obscuration by either the host galaxy or the circumnuclear dust torus might affect the equivalent widths. Given the small observed spatial scales from which most of the line emission emanates, it is unclear how these observations fit into the favored ``disappearing NLR'' scenario for the narrow-line Baldwin effect.

  }

   \maketitle

%

\section{Introduction}\label{Sec:Intro}

The Baldwin effect, a negative correlation between equivalent width, $\EW$, and continuum luminosity, $L$, is commonly found in broad UV/optical emission lines of AGN. Initially, it was reported for the \ion{C}{iv}($\lambda$1549) line \citep{Bal77} for which $\EW\propto L^{-0.2}$, but later also found in emission lines of many other species for which the strength of the correlation depends on the ionization potential of the respective line \citep[e.g.][]{Die02}. Although the Baldwin effect is well established, its physical origin is still a matter of debate. A common explanation refers to the spectral shape of the ionizing continuum which appears softer in high luminosity sources than in AGN with lower luminosities. This results in a relatively lower number of ionizing photons in the broad-line region (BLR) which, in turn, leads to a smaller $\EW$. While this explanation is, at least, qualitatively in agreement with the observed dependence on line ionization energy, it is not clear if $L$ is the fundamental parameter that drives the broad-line Baldwin effect (BLBE). Instead, a similar effect has been observed if $L$ is replaced by the black hole mass, $\MBH$, or the Eddington ratio, $\ledd=L/L_\Edd$ \citep[e.g.][]{Wan99,War04}.

In addition to the BLBE, several narrow optical emission lines have been found to show a Baldwin effect with similar order-of-magnitude luminosity-dependence as well \citep[hereafter: oNLBE; e.g.][]{Bro92,McI99,Cro02}. Given the different spatial scales where the broad- and the narrow-lines are emitted, it is generally believed that the driving physical mechanisms for the BLBE and oNLBE are different \citep[e.g.][ and references therein]{Shi07}. A luminosity-dependence of the equivalent width, sometimes referred to as Iwasawa-Taniguchi effect, is also known for the narrow K$\alpha$-line in X-rays \citep[e.g.][]{Iwa93,Bia07}.

In this letter, we present our discovery of a strong Baldwin effect in nearby Seyfert galaxies for the most prominent narrow emission lines that are present in AGN mid-infrared spectra and observable from the ground, in particular [\ion{Ar}{iii}]($\lambda8.99\,\micron$), [\ion{S}{iv}]($\lambda10.51\,\micron$), and [\ion{Ne}{ii}]($\lambda12.81\,\micron$). The Baldwin Effect in these mid-infrared forbidden lines (hereafter: iNLBE) is much stronger than for any other as yet observed. In Sect~\ref{Sec:Obs}, we report on details about the VLT/VISIR observations. In Sect.~\ref{Sec:Bald}, the results are presented, followed by a discussion of our finding in Sect.~\ref{Sec:Disc}. This paper (hereafter: paper I) presents first results on line emission of our larger campaign using VLT/VISIR which is dedicated to mid-infrared spectroscopy of nearby AGN at high spatial resolution. A forthcoming paper (hereafter: paper II) will deal with the mid-infrared continuum emission of AGN. In the following, we will use cosmological parameters $H_o=73\,{\rm km/(s\,Mpc)}$, $\Omega_\Lambda=0.72$, and $\Omega_m=0.24$ \citep{Spe07}.

\section{Observations}\label{Sec:Obs}

We used the mid-infrared imager and spectrograph VISIR located at the 8\,m UT3 telescope of the ESO/Paranal observatory. With VISIR, we observed a sample of 8 nearby AGN of which 5 are Seyfert 2 (S1h, S1i, S2) and 3 are Seyfert 1 (S1.0, S1.5) galaxies. The $8-13\,\micron$ $N$-band spectra were acquired in low spectral resolution mode (R$\sim$300), for which 4 different wavelength settings centered at 8.5, 9.8, 11.4, and 12.4$\,\micron$ have to be combined for each object. With VISIR, the achieved spatial resolution in this wavelength range is $0\farcs25-0\farcs39$, which is a factor $2-4$ smaller than the used slit widths of $0\farcs75-1\farcs0$ making slit losses negligible. The resulting spectra were extracted using the standard VISIR CPL pipeline (V3.0.0) and calibrated by a number of generic standard stars. For NGC~4507, a single standard star observed within the same night has been used for calibration. In addition to the 8 objects, archival VISIR data of NGC~1068 and NGC~253 has been downloaded, extracted, and calibrated in the same way. A more detailed description of the observations will be presented in paper II. NGC~4593 has been observed only in the 8.5 and 9.8$\,\micron$ settings, corresponding to a restframe wavelength coverage of approximately $7.9-10.4\,\micron$. As a result, only [\ion{Ar}{iii}]($\lambda8.99\,\micron$) is seen in the combined spectrum, giving $\EW$ data in this line on a total of 8 objects. For [\ion{S}{iv}]($\lambda10.51\,\micron$), and [\ion{Ne}{ii}]($\lambda12.81\,\micron$), equivalent widths of 7 objects is available.

\begin{figure}
\centering
\includegraphics[angle=0,width=9.0cm]{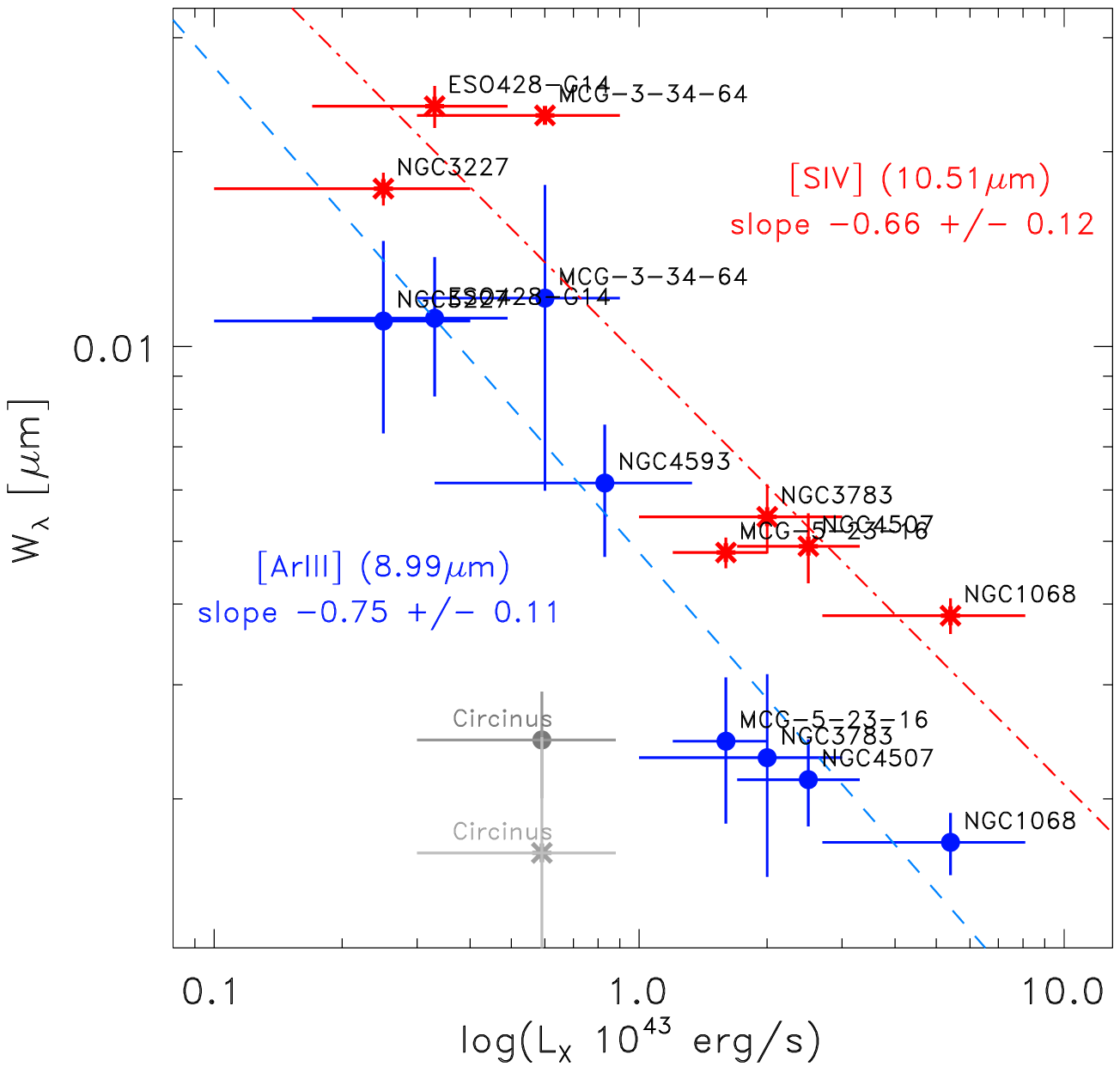}
\caption{Equivalent width, $\EW$, of the [\ion{Ar}{iii}] (blue circles) and [\ion{S}{iv}] (red crosses) lines plotted over the X-ray luminosity, $\LX$, for our sample of nearby AGN as listed in Table~\ref{Tab:EW}. The fitted correlations $\EW \propto \LX^{-0.75\pm0.11}$ for [\ion{Ar}{iii}] (blue dashed line; $\rho_{\rm Spearman}=-0.91$, significance $0.002$), and $\EW \propto \LX^{-0.66\pm0.12}$ for [\ion{S}{iv}] (red dashed-dotted line; $\rho_{\rm Spearman}=-0.79$, significance $0.04$) are shown. Circinus (gray symbols) is outlying probably due to significant obscuration by dust in the host galaxy or the circumnuclear dust torus (see Fig.~\ref{Fig:Obscure}).}\label{Fig:LineComb}
\end{figure}

\begin{figure}
\centering
\includegraphics[angle=0,width=9.0cm]{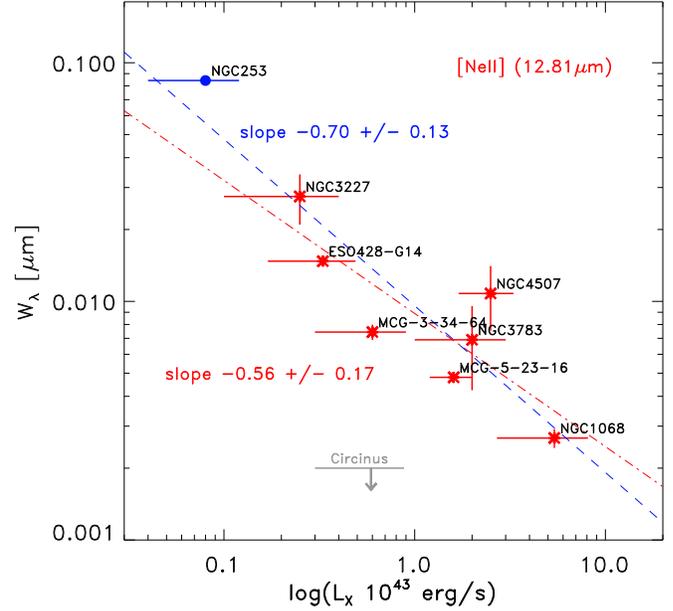}
\caption{Equivalent width, $\EW$, of the [\ion{Ne}{ii}] line plotted over the X-ray luminosity, $\LX$ for 7 nearby Seyfert galaxies (red crosses). The fitted correlation $\EW \propto \LX^{-0.56\pm0.17}$ ($\rho_{\rm Spearman}=-0.75$, significance $0.05$), is shown as red, dashed-dotted line. When including NGC~253 (blue circle) -- despite its questionable AGN nature -- a correlation $\EW \propto \LX^{-0.70\pm0.13}$ ($\rho_{\rm Spearman}=-0.83$, significance $0.01$) is found (blue dashed line). An upper limit for Circinus (gray arrow) is given. The non-detection of [\ion{Ne}{ii}] from the nuclear point source is probably due to significant obscuration by dust in the host galaxy or the circumnuclear dust torus (see Fig.~\ref{Fig:Obscure}).}\label{Fig:NeII}
\end{figure}

\section{The Baldwin effect in forbidden mid-IR lines}\label{Sec:Bald}

\subsection{Results}

\begin{table*}
\centering
\caption{Equivalent widths and X-ray luminosities of our sample of nearby AGN.}\label{Tab:EW}
\begin{tabular}{l c c c c c c c c}
\hline\hline
object & type\,$^a$ & $D_L$\,$^b$ & $10\,\micron$ scale\,$^c$ & $\EW$([\ion{Ar}{iii}]) &  $\EW$([\ion{S}{iv}]) &  $\EW$([\ion{Ne}{ii}]) & $\LX$ & ref. \\
& & (Mpc) & (pc) & ($10^{-4}\,\micron$) & ($10^{-4}\,\micron$) & ($10^{-4}\,\micron$) & ($10^{43}$\,erg/s) & \\ \hline
NGC~1068 & S1h  & 14.4\,$^d$ & 21.4 & $17.13\pm0.94$  &          $38.37\pm1.21$ &          $26.74\pm1.21$  &          $5.4\pm2.7$  & 1 \\
ESO~428--G14 & S2  & 22.3 & 33.2 &   $110.57\pm13.43$ &         $235.16\pm8.74$ &         $147.40\pm2.44$  &          $0.33\pm0.16$ & 1 \\
MCG--5--23--16 & S1i & 39.5 & 58.8 &  $24.55\pm3.12$  &          $48.04\pm1.30$ &         $111.61\pm1.46$  &          $1.7\pm1.0$ & 2  \\
NGC~4507 & S1h  & 54.5 & 80.9 &       $21.41\pm1.64$  &          $49.14\pm3.03$ &         $107.91\pm16.32$ &          $2.5\pm0.8$  & 3  \\
Circinus & S1h  & 4.2\,$^e$ &  6.2 &  $24.66\pm2.31$  &          $16.50\pm4.20$ &         $<27$            &          $0.59\pm0.29$ & 1 \\
MCG--3--34--64 & S1h & 63.3 & 94.2 & $118.78\pm29.46$ &         $227.63\pm1.28$ &          $74.35\pm2.77$  &          $0.6\pm0.3$ & 3  \\
NGC~3783 & S1.5 & 44.7 & 66.5 &       $23.15\pm4.00$  &          $54.56\pm3.33$ &          $68.99\pm13.28$ &          $2.0\pm1.0$ & 3  \\
NGC~3227 & S1.5 & 20.4 & 30.4 &      $109.48\pm18.02$ &         $175.37\pm5.09$ &         $275.10\pm32.69$ &          $0.25\pm0.15$ & 4 \\
NGC~4593 & S1.0 & 42.0 & 62.5 &       $61.52\pm7.19$  &           $\ldots$      &           $\ldots$       &          $0.83\pm0.50$ & 5 \\ 
{\it NGC~253} & {\it S} & {\it 3.5\,$^f$} & {\it 5.2} &     $\mathit{\ldots}$     &     $\mathit{\ldots}$      &     $\mathit{842.27\pm11.13}$ &  $\mathit{0.08\pm0.04}$ & {\it 1} \\  
\hline
\end{tabular}
\begin{list}{}{}
\item --- {\it Notes:} $^a$ AGN types from \citet{Ver06}; $^b$ Luminosity distance based on CMB reference frame redshifts from NED, otherwise indicated; $^c$ Diffraction limit for a 8.2\,m VLT telescope $\sim$0$\farcs$25; $^d$ \citet{Tul88}; $^e$ \citet{Fre77}; $^f$ \citet{Rek05}
\item --- {\it References:} (1) $\LMIR-\LX$-correlation from \citet{Hor08} using 12$\,\micron$ VISIR fluxes for NGC~1068 (14\,Jy), Circinus (10\,Jy), ESO~428--G14 (0.22\,Jy), and {\it NGC~253 (2.1\,Jy)}); (2) \citet{Bal04}; (3) \citet{Hor08}; (4) mean of recent 2--10\,keV \citep{Cap06} and 17--60\,keV \citep{Saz07} data taken, 60\% error inferred due to variability; (5) \citet{Shi06}
\end{list}
\end{table*}

In Table~\ref{Tab:EW}, we present extracted $\EW$ of all three lines. For all objects, the line emission was strongly peaked towards the nucleus. Except for the emission in the most nearby objects NGC~1068 ($\sim10$\% extension in [\ion{S}{iv}] with respect to the continuum PSF) and Circinus ($\sim$50\% extension in [\ion{S}{iv}] and [\ion{Ne}{ii}]), no spatial extension was detected in the continuum {\it and} the lines. The reduction was done in a way that even for these extended objects, almost no line flux is lost by the extraction window. 
The errors on $\EW$ represent the uncertainty due to the determination of the shape of continuum emission underlying the emission line. This uncertainty is in part due to the presence of sky lines in those spectra which were taken under unfavorable atmospheric conditions (e.g. high amount of precipitable water vapor). This affects the [\ion{Ne}{ii}] line the most since strong atmospheric bands are present between $12.4-12.7\,\micron$. On the other hand, the error bars of the [\ion{Ar}{iii}] line are predominantly caused by the general weakness of the line with respect to the continuum.

In addition to the line $\EW$\'s, Table~\ref{Tab:EW} lists (absorption-corrected) $2-10$\,keV X-ray luminosities, $\LX$, for our sample of AGN. These are taken as a measure for the ionizing continuum luminosity since the UV/optical AGN emission of our type 2 sources is heavily affected by dust absorption from the putative circumnuclear dust torus. Unfortunately, 3 of our objects are Compton-thick to the X-ray emission (NGC~1068, Circinus, and ESO~428--G14). For these objects, we estimated the nuclear X-ray luminosity by extracting the 12$\,\micron$ continuum emission from our spectra, which will be presented in paper II, and translating the corresponding spectral luminosity into $\LX$ using the most recent version of the $\LMIR-\LX$-correlation for AGN as established from VISIR observations with similar spatial resolution \citep{Hor08}. We adopt 50\% of error on $\LX$ for these objects.

\begin{figure}
\centering
\includegraphics[angle=0,width=9.3cm]{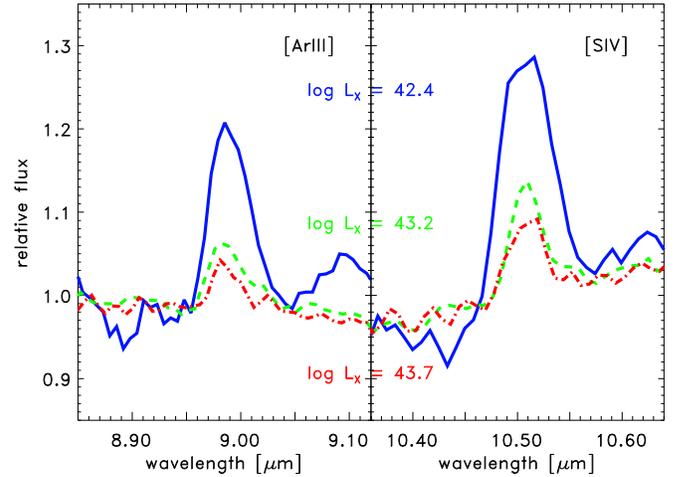}
\caption{[\ion{Ar}{iii}] (left) and [\ion{S}{iv}] (right) spectral region of NGC~3227 ($\log \LX = 42.4$; solid blue line), MCG--5--23--16 ($\log \LX = 43.2$; dashed green line), and NGC~1068 ($\log \LX = 43.7$; dash-dotted red line) illustrating the decrease of $\EW$ with X-ray luminosity. The spectra have been smoothed to flatten the continuum.}\label{Fig:linespec}
\end{figure}

In Figs.~\ref{Fig:LineComb} \& \ref{Fig:NeII}, we show the measured $\EW$ of [\ion{Ar}{iii}], [\ion{S}{iv}], and [\ion{Ne}{ii}] versus $\LX$. The trend of smaller $\EW$ with luminosity is evident. As an illustration, the spectral regions around the  [\ion{Ar}{iii}] and [\ion{S}{iv}] lines are presented in Fig.~\ref{Fig:linespec} for NGC~3227 ($\log \LX = 42.4$), MCG--5--23--16 ($\log \LX = 43.2$), and NGC~1068 ($\log \LX = 43.7$). A statistical analysis of the correlation for our sample shows Spearman ranks $\rho_{\rm Spearman} = -0.75\ldots-0.9$ with significance levels in the range of $0.2-5\times10^{-2}$. Despite the limited number of objects and luminosity coverage, correlations evolving out of the data can be confidently established. For BLBE and oNLBE studies, higher significance is achieved by averaging spectra of different objects within a narrow luminosity range. This overcomes peculiarities of individual objects \citep[e.g.][]{Cro02}. We aim for observations of a larger sample to make similar studies for the presented iNLBE. It is important to note that no correlation of $\EW$ with AGN distance, $D_L$, is present in our sample ($\rho_{\rm Spearman}<0.25$, significance $>0.6$ for all lines).

As of yet, only one other study mentions a possible Baldwin effect in the mid-infrared. In an AAS abstract, \citet{Ker06} presented the detection of a Baldwin effect for the [\ion{S}{iv}] line in AGN data obtained by the {\it Spitzer} satellite. They also note indications of a Baldwin effect for [\ion{Ne}{ii}], admitting that their study is suffering from the low spatial resolution of the {\it Spitzer} data that they used. In the available abstract, nothing is mentioned about a slope or the scatter of the anti-correlation. Here, we demonstrate that the Baldwin effect of both the [\ion{S}{iv}] and [\ion{Ne}{ii}] line, and in addition the [\ion{Ar}{iii}] line, is quite significant -- and strong -- when using high spatial resolution data, even with a small object sample. Thus, high spatial resolution appears to be crucial.


\subsection{Relation to $\MBH$ and $\ledd$}\label{Sec:MBH}

For the three type 1 AGN in our sample, black hole masses, $\MBH$, have been estimated based on reverberation mapping data. We adopt black hole masses for NGC~3227 ($\log\MBH(\Msun) = 7.63 \pm 0.31$) and NGC~3783 ($\log\MBH = 7.47 \pm 0.08$) from \citet{Onk04}, and $\log\MBH = 6.99 \pm 0.10$ for NGC~4593 from \citet{Den06}. In addition, the black hole mass of the Seyfert 2 galaxy NGC~1068 could be determined by MASER cloud kinematics as $\log\MBH \approx 7.0$ \citep{Gre96}. For these four AGN, we analyzed the dependence of the equivalent widths of the [\ion{Ar}{iii}] line on $\MBH$. A statistical test shows no evident correlation for our limited sample of objects and the small coverage of black hole masses ($\rho_{\rm Spearman}=0.40$, significance $0.6$, for a nominal fit $\EW$([\ion{Ar}{iii}])$\propto \MBH^{0.5\pm0.7}$). Subject to the limitations, this indicates a fundamental difference of the iNLBE as compared to the BLBE for which such an anti-correlation has been found \citep{War04}.

By using bolometric correction to the X-ray luminosities listed in Table~\ref{Tab:EW} \citep{Mar04}, it is possible to estimate the Eddington ratio, $\ledd$, of four AGN with known $\MBH$. A statistical test for a correlation between $\EW($[\ion{Ar}{iii}]$)$ and $\ledd$ 
reveals a Spearman rank $\rho_{\rm Spearman} = -0.8$ with a significance of 0.2 for the relation $\EW$([\ion{Ar}{iii}])$\propto \ledd^{-0.40\pm0.17}$.
Thus, our limited sample doesn't allow us to firmly establish a correlation, but a negative dependence of $\EW$ on $\ledd$ is indicated.


\subsection{The case of Circinus}

We excluded Circinus from all analysis of the correlations. As can be seen in Figs.~\ref{Fig:LineComb} \& \ref{Fig:NeII}, the $\EW$'s of all three mid-infrared lines deviate significantly from the nominal fit to the other AGN. Although it could be a resolution effect (Circinus is the nearest AGN in the sample), NGC~1068 has about the same spatial resolution when scaled for the different luminosities (scaling $r\propto L^{1/2}$). Moreover, we did not detect extended emission in the lines down to $\sim$5\% of the peak flux. This gives us some confidence to reject that a significant amount of line flux is over-resolved and lost in the noise.

\begin{figure}
\centering
\includegraphics[angle=0,width=8.8cm]{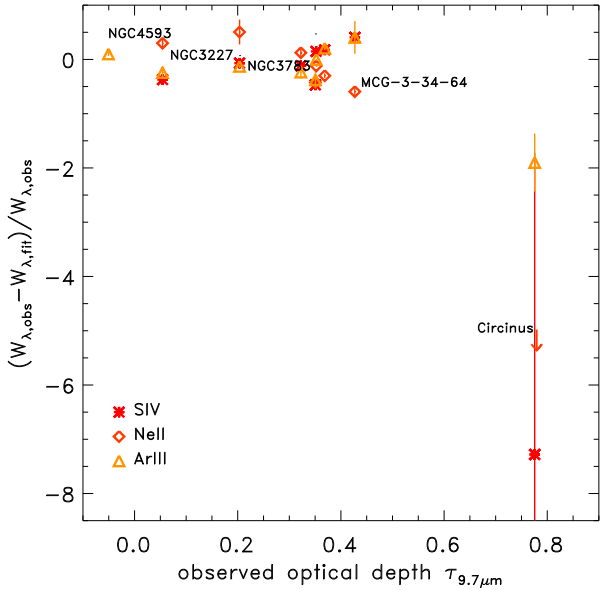}
\caption{Relative deviation of the observed $\EW$ from the correlation for all three lines plotted against the observed optical depth, $\tau_{\rm 9.7\,\micron}$, in the silicate feature. Circinus, which is an outlier in the Baldwin correlations of all three lines, is the object with the deepest silicate feature. Due to radiative transfer effects, the actual obscuration of the central AGN by dust from the putative torus might be even stronger than suggested by the $\tau_{\rm 9.7\,\micron}$ value.}\label{Fig:Obscure}
\end{figure}

Circinus is known for strong dust extinction towards the nucleus by dust lanes in both the host and our own Galaxy. In Fig.~\ref{Fig:Obscure}, we plot the deviation from the nominal correlation with X-ray luminosity (as given in Figs.~\ref{Fig:LineComb} \& \ref{Fig:NeII}) of each line against the depth of the silicate feature. Circinus shows the strongest deviations from all correlations and, at the same time, has the deepest silicate feature. It is deeper than in any other type 2 AGN in our sample. In case of foreground extinction (host or our galaxy), we expect that the flux within the lines is reduced in the same way without a change in $\EW$. On the other hand, if part or all of the line emission is originating from inward of the torus, the extinction in the lines could be much stronger in a more or less edge-on torus geometry. 
Given the fact that the bulk of line emission in all other AGN is coming from the unresolved point source, such a scale and inclination effect appears possible. Strong extinction towards the NLR based on mid-infrared spectra of Circinus was also noted by \citet{Roc06}.

\section{Discussion and conclusions}\label{Sec:Disc}

\subsection{A comparison to UV/optical lines}\label{Sec:Comp}

The negative correlations in the [\ion{Ar}{iii}], [\ion{S}{iv}], and [\ion{Ne}{ii}] lines as observed with VLT/VISIR is much stronger than any known Baldwin effect in optical/UV broad or narrow lines. This is remarkable since the spatial resolution in the optical of the most recent studies should be similar or even better than what is achieved in the mid-infrared with the VLT. In UV/optical lines, a trend for larger correlation slopes with increasing ionization potential is observed \citep[e.g.][]{Die02}. The potential of the presented lines is in the range of 40--47\,eV and, thus, comparable to some of the optical/UV lines. From the optical/UV lines, slopes of the order of $-0.1$ to $-0.2$ would be expected. The measured slope of the mid-infrared lines deviates by at least 4-$\sigma$ from that expectation (except [\ion{Ne}{ii}] w/o NGC~253: 2-$\sigma$).

In Sect~\ref{Sec:MBH}, we found some negative trend of $\EW$([\ion{Ar}{iii}]) with $\ledd$, but possibly not with $\MBH$. If confirmed, this is in contrast to the observed and theoretically-expected behavior of the broad lines \citep[e.g.][]{Wan99,War04}. Thus, the unsettled correlation with $\MBH$ might imply that the iNLBE has a different physical origin than the BLBE, in particular questioning the relevance of the $\MBH$-dependence of the ionising continuum on the mid-infrared narrow-line emission.

\subsection{Implications for the origin of the iNLBE}

One suggestion for the origin of the oNLBE relates to the size scaling of the narrow-line region (NLR), for which a luminosity dependence $\RNLR \propto L^{0.5}$ has been found \citep{Ben02}. As a consequence, the NLR of luminous sources may reach galactic size scales and, thus, lose its gas. This scenario is often referred to as the ``disappearing NLR'' \citep{Cro02,Net06}. For the mid-infrared lines, we find that the main portion of their emission is not coming from a spatially-extended region, e.g. as for [\ion{O}{iii}] in the optical, but from scales smaller than 100\,pc. If the lack of line emission in Circinus is really an inclination-dependent obscuration effect, then the involved scales are significantly smaller. Since it is difficult to imagine how the NLR could disappear on these scales, the iNLBE probably requires a different explanation. This explanation needs to explain, in particular, (1) the steepness of the correlation of the presented lines, and (2) the fact that the mid-infrared lines do not follow the ionisation potential-dependence as observed in optical lines. Possible mechanisms might involve a luminosity-dependent increase of density in the inner NLR (thus surpressing forbidden line emission).

\begin{acknowledgements}
We thank L. Spinoglio for helpful comments and suggestions which improved the paper. This research made use of the NASA/IPAC Extragalactic Database (NED) operated by the JPL (Caltech), under contract with NASA. PG is supported by JSPS and RIKEN Foreign postdoctoral fellowships. HH acknowledges support by DFG via SFB 439.
\end{acknowledgements}

\bibliographystyle{aa}

\end{document}